\input harvmac
\input amssym.def
\input amssym.tex

\def\AdS{$AdS$}
\def\CFT{$CFT$}
\def\BPS{$BPS$}
\def\bk{{\bf N}}
\def\bkb{{\bf\overline N}}
\def\H{{\rm H}}
\def\^{{\wedge}}

\def\BC{{\Bbb C}}
\def\BP{{\Bbb P}}
\def\BR{{\Bbb R}}
\def\BZ{{\Bbb Z}}
\def\CB{{\cal B}}
\def\CC{{\cal C}}
\def\CM{{\cal M}}
\def\CN{{\cal N}}
\def\CO{{\cal O}}
\def\CR{{\cal R}}

\noblackbox

\def\urlfont{\hyphenpenalty=10000 \hyphenchar\tentt='057 \tt}

\newbox\tmpbox\setbox\tmpbox\hbox{\abstractfont PUPT-2044}
\Title{\vbox{\baselineskip12pt\hbox{\hss hep-th/0207125}
\hbox{PUPT-2044}}}
{\vbox{
\centerline{BPS Branes From Baryons}}}
\smallskip
\centerline{Chris E. Beasley}
\smallskip
\centerline{\it{Joseph Henry Laboratories, Princeton University}}
\centerline{\it{Princeton, New Jersey 08544}}
\bigskip\bigskip
We elucidate the relationship between supersymmetric $D3$-branes and
chiral baryonic operators in the \AdS/\CFT\ correspondence.  For
supersymmetric backgrounds of the form $AdS_5 \times H$, we
characterize via holomorphy a large family of supersymmetric $D3$-brane 
probes wrapped on $H$.  We then quantize this classical family of
probe solutions to obtain a \BPS\ spectrum which describes 
$D3$-brane configurations on $H$.  For the particular examples $H =
T^{1,1}$ and $H = S^5$, we match the \BPS\ spectrum to the spectrum of 
chiral baryonic operators in the dual gauge theory.
  
\Date{July 2002}

\lref\adscft{See O. Aharony, S.~S. Gubser, J. Maldacena, H. Ooguri, and
Y. Oz, ``Large N Field Theories, String Theory, and Gravity,''
Phys.Rept. {\bf 323} (2000) 183--386, {\urlfont hep-th/9905111}.}

\lref\aps{M. Aganagic, C. Popescu, and J.~H. Schwarz, ``D-Brane
Actions with Local Kappa Symmetry,'' Phys. Lett. {\bf B393} (1997) 311--315, 
{\urlfont hep-th/9610249}.}

\lref\apsa{M. Aganagic, C. Popescu, and J.~H. Schwarz,
``Gauge-Invariant and Gauge-Fixed D-brane Actions,'' Nucl. Phys. {\bf
B495} (1997) 99--126, {\urlfont hep-th/9612080}.}

\lref\bar{C. B\"ar, ``Real Killing Spinors and Holonomy,''
Comm. Math. Phys. {\bf 154} (1993) 509--521.}

\lref\bbns{V. Balasubramanian, M. Berkooz, A. Naqvi, and M. Strassler,
``Giant Gravitons in Conformal Field Theory,'' JHEP {\bf 0204} (2002)
034, {\urlfont hep-th/0107119}.}

\lref\bhln{V. Balasubramanian, M. Huang, T. Levi, and A. Naqvi, ``Open
Strings From $\CN=4$ Super Yang Mills,'' {\urlfont hep-th/0204196}.}

\lref\bbs{K. Becker, M. Becker, and A. Strominger, ``Fivebranes,
Membranes, and Non-Perturbative String Theory,'' Nucl. Phys. {\bf
B456} (1995) 130--152, {\urlfont hep-th/9507158}.}

\lref\bhk{D. Berenstein, C.~P. Herzog, and I.~R. Klebanov, ``Baryon
Spectra and \AdS/\CFT\ Correspondence'', JHEP {\bf 0206} (2002) 047, 
{\urlfont hep-th/0202150}.}

\lref\bg{C. Boyer and K. Galicki, ``On Sasakian-Einstein Geometry,''
Int. J. Math. {\bf 11} (2000) 873--909, {\urlfont math.DG/9811098}.}

\lref\bkop{E. Bergshoeff, R. Kallosh, T. Ort\'in, and G. Papadopoulos,
``$\kappa$-Symmetry, Supersymmetry, and Intersecting Branes,''
Nucl. Phys. {\bf B502} (1997) 149--169, {\urlfont hep-th/9705040}.}

\lref\blaui{M. Blau, J. Figueroa-O'Farrill, C. Hull, and
G. Papadopoulos, ``A New Maximally Supersymmetric Background of IIB
Superstring Theory,'' JHEP {\bf 0201} (2002) 047, {\urlfont
hep-th/0110242}.} 

\lref\blauii{M. Blau, J. Figueroa-O'Farrill, C. Hull, and
G. Papadopoulos, ``Penrose Limits and Maximal Supersymmetry,''
Class. Qaunt. Grav. {\bf 19} (2002) L87--L95, {\urlfont hep-th/0201081}.}

\lref\bmn{D. Berenstein, J. Maldacena, and H. Nastase, ``Strings in
Flat Space and PP Waves from $\CN=4$ Super Yang Mills,'' JHEP {\bf
0204} (2002) 013, {\urlfont hep-th/0202021}.}

\lref\bt{E. Bergshoeff and P.~K. Townsend, ``Super D-branes,''
Nucl. Phys. {\bf B490} (1997) 145--162, {\urlfont hep-th/9611173}.}

\lref\candelas{P. Candelas and X. de la Ossa, ``Comments on
Conifolds,'' Nucl. Phys. {\bf B342} (1990) 246--268.}

\lref\cgnw{M. Cederwall, A. von Gussich, B. Nilsson, and
A. Westerberg, ``The Dirichlet Super-Three-Brane in Ten-Dimensional
Type IIB Supergravity,'' Nucl. Phys. {\bf B490} (1997) 163--178, 
{\urlfont hep-th/9610148}.}

\lref\cgnsw{M. Cederwall, A. von Gussich, B.~E.~W. Nilsson,
P. Sundell, and A. Westerberg, ``The Dirichlet Super p-Branes in Ten
Dimensional Type IIA and IIB Supergravity,'' Nucl. Phys. {\bf B490}
(1997) 179--201, {\urlfont hep-th/9611159}.}

\lref\cox{D. Cox, ``The Homogeneous Coordinate Ring of a Toric
Variety,'' J. Algebraic Geometry {\bf 4} (1995) 17--50, {\urlfont
math.AG/9210008}.}

\lref\fk{Th. Friedrich and I. Kath, ``Einstein Manifolds of Dimension
Five with Small First Eigenvalue of the Dirac Operator,''
J. Diff. Geom. {\bf 29} (1989) 263--279.}

\lref\fulton{W. Fulton, {\it Introduction to Toric Varieties},
(Princeton Univ. Press, 1993).}

\lref\gmt{M.~T. Grisaru, R.~C. Myers, and \O. Tafjord, ``SUSY and
Goliath'', JHEP {\bf 0008} (2000) 040, {\urlfont hep-th/0008015}.}

\lref\gk{S.~S. Gubser and I.~R. Klebanov, ``Baryons and Domain Walls
in an $\CN=1$ Superconformal Gauge Theory,'' Phys. Rev. {\bf D58}
(1998) 125025, {\urlfont hep-th/9808075}.}

\lref\grw{S. Gukov, M. Rangamani, and E. Witten, ``Dibaryons, Strings,
and Branes in \AdS\ Orbifold Models,'' JHEP {\bf 9812} (1998) 025,
{\urlfont hep-th/9811048}.}

\lref\gut{J. Gutowski and G. Papadopoulos, ``AdS Calibrations,''
Phys. Lett. {\bf B462} (1999) 81--88, {\urlfont hep-th/9902034}.}

\lref\harvey{R. Harvey and H.~B. Lawson, Jr., ``Calibrated
Geometries,'' Acta Math. {\bf 148} (1982) 47--157.} 

\lref\hhi{A. Hashimoto, S. Hirano, and N. Itzhaki, ``Large Branes in
AdS and Their Field Theory Dual'', JHEP {\bf 0008} (2000) 051, 
{\urlfont hep-th/0008016}.}

\lref\kw{I.~R. Klebanov and E. Witten, 
``Superconformal Field Theory on Threebranes at a
Calabi-Yau Singularity,'' Nucl. Phys. {\bf B536} (1998) 199--218,
{\urlfont hep-th/9807080}.}

\lref\kehagias{A. Kehagias, ``New Type IIB Vacua and Their F-Theory 
Interpretation,'' Phys. Lett. {\bf B435} (1998) 337--342, 
{\urlfont hep-th/9805131}.}

\lref\kim{H.~J. Kim, L.~J. Romans, and P. van Nieuwenhuizen, ``The
Mass Spectrum of Chiral $\CN = 2$ $D=10$ Supergravity on $S^5$,''
Phys. Rev. {\bf D32} (1985) 389--399.}

\lref\mets{R.~R. Metsaev, ``Type IIB Green-Schwarz Superstring in
Plane Wave Ramond-Ramond Background,'' Nucl. Phys. {\bf B625} (2002)
70--96, {\urlfont hep-th/0112044}.}

\lref\mst{J. McGreevy, L. Susskind, and N. Toumbas, ``Invasion of the
Giant Gravitons from Anti-de Sitter Space,'' JHEP {\bf 0006} (2000)
008, {\urlfont hep-th/0003075}.}

\lref\mik{A. Mikhailov, ``Giant Gravitons from Holomorphic Surfaces,''
JHEP {\bf 0011} (2000) 027, {\urlfont hep-th/0010206}.}

\lref\mp{David R. Morrison and M. Ronen Plesser, 
``Non-Spherical Horizons, I,'' Adv. Theor. Math. Phys. {\bf 3} (1999)
1--81, {\urlfont hep-th/9810201}.}

\lref\ooy{H. Ooguri, Y. Oz, and Z. Yin, ``D-Branes on Calabi-Yau
Spaces and Their Mirrors,'' Nucl. Phys. {\bf B477} (1996) 407--430,
{\urlfont hep-th/9606112}.}

\lref\wbb{E. Witten, ``Baryons and Branes in Anti-de Sitter Space,''
JHEP {\bf 9807} (1998) 006, {\urlfont hep-th/9805112}.}

\lref\woodhouse{N.~M.~J. Woodhouse, {\it Geometric Quantization,
Second Ed.,} (Clarendon Press, Oxford, 1992).}

\newsec{Introduction.}

The \AdS/\CFT\ correspondence \adscft\ is a remarkable example of the
long-suspected relationship between string theories and gauge
theories.  Yet, despite the many compelling pieces of evidence that 
exist for a duality between type-IIB string theory on $AdS_5 \times
S^5$ and the $\CN=4$ super Yang-Mills theory, we still do not
understand such a fundamental issue as how the Yang-Mills theory 
encodes the full spectrum of the string theory.  

One difficulty, of course, is that even the perturbative string
spectrum in the \AdS\ background is not known, owing to the presence
of Ramond-Ramond five-form flux.  Nevertheless, we can circumvent this 
difficulty in various ways.  For instance, the massless modes of the string
just give rise to the spectrum of type-IIB supergravity in the \AdS\
background, which can be directly computed \kim.  These
supergravity states fall into short representations of the
superconformal algebra and can be matched to chiral primary
operators in the Yang-Mills theory.

Another more recent approach to this problem has been to take a
Penrose limit in which the $AdS_5 \times S^5$ background reduces to a
pp-wave background \blaui, \blauii\ for which the classical string spectrum can
be computed exactly in a Green--Schwarz formalism \mets.  In this limit
the perturbative string states can be matched to ``nearly
chiral'' operators which have large charge under a $U(1)$ subgroup of
the global $SU(4)$ $\CR$-symmetry of the Yang-Mills theory \bmn.

In this paper, we approach the problem of how the Yang-Mills theory
describes ``stringy'' objects by studying the non-perturbative sector of 
\BPS\ states which arise from supersymmetric $D3$-branes in the \AdS\ 
background.  Paradoxically, despite the fact that we can't quantize a 
string in this background, our main result is to quantize a
supersymmetric $D3$-brane, which is possible because the brane couples very
simply to the Ramond-Ramond flux.  

Supersymmetry plays an essential role in our work, so we consider a 
general background of the form $AdS_5 \times H$ which preserves at least
eight supercharges.  Here $H$ is a compact 
Einstein five-fold of positive curvature which supports $N$ units 
of five-form flux, 
\eqn\flux{ \int_{H} \, F_5 \; = \; N \,.}
In this background we study supersymmetric $D3$-branes which wrap
three-cycles in $H$ and appear as particles in $AdS_5$.

Such backgrounds have already been studied extensively \kehagias, \kw, \mp\ 
in the context of the \AdS/\CFT\ correspondence.  Nor is the idea of studying 
supersymmetric $D3$-branes wrapped on three-cycles in $H$ new \wbb,
\gk, \grw.  In these, by now classic, references, the 
authors study supersymmetric $D3$-branes wrapped on topologically non-trivial
three-cycles in $H = \BR \BP^5$, $T^{1,1}$, and $S^5 / \BZ_3$.  In
each of these cases, due to the symmetry of $H$, a supersymmetric
$D3$-brane can wrap any of a family of three-manifolds $\Sigma \subset
H$.  When the corresponding collective coordinates of the brane
are quantized, the resulting zero-modes can be matched directly to a 
set of Pfaffian or di-baryonic chiral operators in the dual gauge theories.

In a similar vein, supersymmetric $D3$-branes which wrap topologically
trivial three-cycles in $H$ have also been studied.  The best example
of such a brane is the giant graviton \mst\ on $H = S^5$.  It 
is a $D3$-brane which wraps a topologically trivial $S^3$ but is still 
supersymmetric \gmt, \hhi\ due to the angular momentum it 
carries as it orbits about a transverse circle in $H$.  Largely in
analogy to the di-baryonic operators which describe non-trivially
wrapped branes, the authors of \bbns\ have proposed that the \BPS\ 
states associated to giant gravitons correspond to sub-determinant
operators in the $\CN=4$ Yang-Mills theory.

These examples suggest a general correspondence between the spectrum
of \BPS\ states arising from $D3$-branes wrapped on three-cycles in
$H$ and the spectrum of chiral baryonic operators in the dual gauge 
theory.  However, as insightfully observed by Berenstein, Herzog, and 
Klebanov \bhk, we still have some things to learn before we fully
understand this correspondence.

For instance, as we shall review in Section 2, the $\CN=1$ gauge
theory dual to the $H = T^{1,1}$ background possesses a tower of 
chiral baryonic operators of increasing conformal dimension $\Delta$.  
The baryons at the base of the tower correspond to the zero-modes of a 
supersymmetric $D3$-brane.  But what $D3$-brane states do the baryons 
of higher $\Delta$ describe?  The question of what this baryon
spectrum really ``means'' cuts to the heart of how the gauge theory 
describes the geometry of a brane configuration on $H$.  Yet, 
as we show, this question is tractable, even simple, because of
supersymmetry\foot{We are hopeful that our results in understanding
the correspondence between \BPS\ states of $D3$-branes and chiral 
operators of the gauge theory will be useful in understanding the 
analogous correspondence of non-\BPS\ states and non-chiral operators, 
which in the pp-wave limit on $S^5$ has already been considered in \bhln.}.

The simplest guess would be that these baryons just describe 
multi-particle states consisting of some zero-mode of the $D3$-brane 
plus particles in the supergravity multiplet on $H$.  That guess is
generally wrong.  Although some of the baryons of higher $\Delta$
factor into products of chiral operators consistent with this
hypothesis, most baryons do not factor in any simple way \bhk.  So we
must somehow interpret these baryons as corresponding to new 
irreducible, \BPS\ states of $D3$-branes.

In practical terms, the crux of this problem is to identify the
spectrum of \BPS\ states in the string theory which arise from 
supersymmetric $D3$-branes wrapped on three-cycles in a given 
(possibly trivial) homology class of $H$.  As it turns out, once we
have computed the \BPS\ spectrum, we can trivially match it to the
spectrum of chiral baryons in examples such as $H = T^{1,1}$ or $H =
S^5$ for which the dual gauge theory is known.

One approach to computing the above \BPS\ spectrum is to use the 
Dirac-Born-Infeld action for a probe $D3$-brane.  The $DBI$
action is appropriate for describing a weakly-curved $D$-brane at 
weak string coupling.  Hence this description of the $D3$-brane is
appropriate in the supergravity regime of a weakly-curved background, 
corresponding to large 't Hooft coupling in the gauge theory.
Provided the Einstein metric on $H$ is known, one can directly use the $DBI$
action to compute perturbatively the spectrum of small fluctuations 
about a particular supersymmetric $D3$-brane configuration in $H$.  

The authors of \bhk\ performed this computation for the well-known
supersymmetric $D3$-branes in $H = T^{1,1}$ and, amidst a plethora of 
modes, they identified a subset of \BPS-saturated states having 
quantum numbers consistent with some of the chiral baryonic operators 
of higher $\Delta$.  Hence they provided compelling evidence for the 
general idea that chiral baryonic operators of higher $\Delta$ 
should correspond to additional \BPS\ states of wrapped $D3$-branes.

We provide a complementary analysis of the \BPS\ spectrum.  In any
quantum mechanics problem, a thorough understanding of the space $\CM$
of classical solutions is often an essential ingredient in quantization.
So rather than starting from a particular supersymmetric $D3$-brane 
and studying its small fluctuations, we begin in Section 3 by asking,
``How can we characterize the general supersymmetric $D3$-brane
configuration on $H$?''  Luckily, Mikhailov \mik\ has already provided
such a characterization via holomorphy, and we can apply his results directly 
to describe $\CM$.  In the example $H = T^{1,1}$, we explicitly
present many new supersymmetric $D3$-brane configurations on $H$, all 
of which are generalizations of the giant graviton solution on $S^5$
to non-trivially wrapped branes on $H$.

In Section 4, we quantize the classical, supersymmetric $D3$-brane
configurations which we found in Section 3.  That is, we determine the
Hilbert space of \BPS\ states associated to $\CM$ as well as the
representations in which these states transform under the 
various global symmetries.  This exercise is a direct 
generalization of the quantization of collective coordinates discussed
in \wbb, \gk, and \grw.  We discuss the examples $H = T^{1,1}$ and $H
= S^5$ in detail.  As a result, we find an immediate correspondence between
\BPS\ states and chiral baryonic operators in the dual gauge
theories.  We finally discuss some issues relating to multi-brane 
states in string theory and factorizable baryonic operators in gauge theory.

\newsec{$T^{1,1}$, the conifold, baryons, and all that.}

Although much of the discussion in Sections 3 and 4 holds for a general
$H$ which preserves supersymmetry, the example $H = T^{1,1}$ is very
good to keep in mind.  In order that this paper be self-contained and
to establish notation, we now review some aspects of the
correspondence between type-IIB string theory on $AdS_5 \times
T^{1,1}$ and its dual $\CN=1$ gauge theory \kehagias, \kw, \mp.  We 
also review the relation, which we generalize in Section 4, between 
wrapped $D3$-branes and baryonic operators proposed in \gk.  We
finally describe the observations of \bhk\ which inspired our work.
So in this section, $H = T^{1,1}$.

\subsec{The geometry of $T^{1,1}$ and the conifold.}

Recall that $H$ is the homogeneous space $SU(2) \times SU(2) /
U(1)$, where the $U(1)$ is a diagonal subgroup inside a maximal
torus of $SU(2) \times SU(2)$.  The manifold underlying $SU(2)$
is $S^3$, and the $S^1$ fiber of the Hopf fibration $S^1
\rightarrow S^3 \rightarrow S^2$ corresponds to a maximal torus of
$SU(2)$.  So the quotient $SU(2) \times SU(2) / U(1)$ can also be 
described as an $S^1$-fibration over $S^2 \times S^2$; equivalently
$H$ is a principal $U(1)$-bundle over $S^2 \times S^2$.

This five-manifold $H$ possesses a unique Einstein metric 
for which the $AdS_5 \times H$ background preserves 
eight supercharges.  This Einstein metric is naturally written in 
coordinates adapted to the $S^1$-fibration.  Let  
$(\theta_1, \phi_1)$ and $(\theta_2,\phi_2)$ be the usual angular coordinates 
on the two-spheres in the $S^2 \times
S^2$ base, and let $\psi \in [0, 4 \pi)$ be an angular coordinate on
the $S^1$ fiber.  Then the Einstein metric is 
\eqn\mtii{ ds_H^2 = {1 \over 9} \left( d \psi + \sum_{i=1}^2
\cos{\theta_i} \, d \phi_i \right)^2 + {1 \over 6} \sum_{i=1}^2 \left( d 
\theta_i^2 + \sin^2{\theta_i} \, d \phi_i^2 \right) \,,}
which satisfies
\eqn\norms{ R_{mn} = 4 \, g_{mn} \,.}
An important property of this metric is that it possesses an
$SU(2) \times SU(2) \times U(1)$ group of isometries.  The $U(1)$
isometry is generated by ${\partial \over {\partial \psi}}$, and the 
$SU(2) \times SU(2)$ isometries act on the $S^2 \times S^2$ base,
which we can regard as the K\"ahler-Einstein manifold $\BC \BP^1 \times \BC
\BP^1$.

Topologically, $H$ is equivalent to $S^2 \times S^3$, so that
$\H_3(H) = \BZ$.  $H$ has two obvious families of three-spheres,
which arise as the loci of fixed $(\theta_1, \phi_1)$ or fixed 
$(\theta_2, \phi_2)$ on the $S^2 \times S^2$ base.  We denote the 
corresponding classes in $\H_3$ by $A$ and $B$.  Both $A$ and $B$
generate $\H_3$, and since the members of the respective families are 
oppositely oriented, $A = - B$.

The superconformal theory dual to the $AdS_5 \times H$ background arises
from the infrared limit of the worldvolume theory on coincident
$D3$-branes at the conifold singularity $X$.  Recall that $X$ can be 
presented algebraically as the locus of points $(z_1, z_2, z_3, z_4) \in \BC^4$
which satisfy
\eqn\con{ z_1 z_2 - z_3 z_4 = 0\,.}
This locus is preserved by the $\BC^\times$-action
\eqn\scale{ (z_1,z_2,z_3,z_4) \rightarrow (\lambda z_1, \lambda z_2, 
\lambda z_3, \lambda z_4)\,,\quad \lambda \in \BC^\times \,.}
When $\lambda$ is real, the $\BC^\times$-action amounts to 
scaling in $\BC^4$, so we see that $X$ is a cone.  This
$\BC^\times$-action is free away from the origin of $\BC^4$, where $X$ 
possesses an isolated conical singularity.

$X$ is actually a Calabi-Yau cone, and a Ricci-flat K\"ahler metric on
$X$ can be determined very simply using symmetries.  By a linear
change of variables in $\BC^4$, we can re-write the defining equation
\con\ of $X$ as
\eqn\conb{ z_1^2 + z_2^2 + z_3^2 + z_4^2 = 0 \,.}
From \conb, we see that $SO(4)$ acts in a natural way on $X$, and we
look for a K\"ahler metric with this group 
acting by isometries.  The associated K\"ahler potential $K$ can also
be chosen to be invariant under $SO(4)$ and hence must be a
function only of 
\eqn\norm{ |z|^2 \equiv |z_1|^2 + |z_2|^2 + |z_3|^2 + |z_4|^2\,.}

Because $|z|^2$ is invariant under the $U(1)$ part of the
$\BC^\times$-action in \scale, under which
\eqn\uoner{ (z_1,z_2,z_3,z_4) \rightarrow (e^{i \alpha} z_1, e^{i
\alpha} z_2, e^{i \alpha} z_3, e^{i \alpha} z_4)\,,\quad \alpha \in
[0,2 \pi) \,,}
the K\"ahler metric actually has an $SO(4) \times U(1)$ group of
isometries.  As the holomorphic three-form $\Omega$ of $X$,
\eqn\omg{ \Omega = {{dz_2 \, \^ \, dz_3 \, \^ \, dz_4} \over
{z_1}}\,,}
has charge two under \uoner, we identify this $U(1)$ isometry with an
$\CR$-symmetry.  

If the K\"ahler metric is to be conical, $K$ must transform 
homogeneously under the scaling with real $\lambda$ in \scale.  Thus, 
$K = |z|^{2 \gamma}$ for some exponent $\gamma$.  To fix $\gamma$, we
note that $\Omega$ transforms as $\Omega \rightarrow \lambda^2 \,
\Omega$ under \scale.   Under the same scaling, the K\"ahler form $\omega$,  
\eqn\kah{ \omega = -i \, {{\partial^2 K} \over {\partial z_i \, \partial 
\overline{z}_j}} \, dz_i \^ d\overline{z}_j \,,}
transforms as $\omega \rightarrow \lambda^{2 \gamma} \, \omega$.  The 
Calabi-Yau condition finally implies that 
\eqn\cy{ \omega \^ \omega \^ \omega = -i \, \Omega \^
\overline{\Omega}\,,} 
which determines $\gamma = 2 / 3$.

Let us return to the defining equation \con\ of $X$.  We can solve
this equation by introducing four homogeneous coordinates 
$A_1$, $A_2$, $B_1$, $B_2$, and setting 
\eqn\zs{ z_1 = A_1 B_1 \,, \quad z_2 = A_2 B_2 \,, \quad z_3 = A_1 B_2
\,, \quad z_4 = A_2 B_1 \,.}  The coordinates $(A_1, A_2, B_1, B_2)$
are homogeneous (as opposed to true) coordinates on $X$ because the
affine coordinates $(z_1,z_2,z_3,z_4)$ of $\BC^4$ are left invariant
under the $\BC^\times$-action,
\eqn\quot{(A_1,\; A_2,\; B_1,\; B_2) \rightarrow 
(\mu A_1,\; \mu A_2,\; \mu^{-1} B_1,\; \mu^{-1} B_2)\,,\quad \mu \in
\BC^\times \,.}
Away from the singularity at the origin, we can fix the real part of this 
action by requiring that
\eqn\moment{ |A_1|^2 + |A_2|^2 = |B_1|^2 + |B_2|^2 \,.}  
We are left with the $U(1)$-action given by
\eqn\uone{ (A_1,A_2,B_1,B_2) \rightarrow (e^{i \beta} A_1, e^{i
\beta} A_2, e^{-i \beta} B_1, e^{-i \beta} B_2)\,,\quad \beta \in
[0,2 \pi)\,.}
In terms of the homogeneous coordinates, $X$ is thus obtained 
by restricting to the locus given by \moment\ and dividing by the residual 
$U(1)$-action of \uone.

We remark that the introduction of homogeneous coordinates on $X$ is 
much less ad hoc than it may appear.  $X$ is a toric variety \fulton, 
and toric varieties are generalizations of projective and 
weighted projective spaces.  As such, any toric variety is endowed
with a set of global homogeneous coordinates \cox\ which function much 
like the usual projective coordinates of projective space.

The five-manifold $H$ can be obtained from $X$ by further restricting
to the locus 
\eqn\hor{ |A_1|^2 + |A_2|^2 = |B_1|^2 + |B_2|^2 = 1 \,.}  
This restriction amounts to a quotient by the scaling part of the
$\BC^\times$-action \scale\ on $X$, after the fixed point at the
origin is excised.  To see that the description of $H$ using the
homogeneous coordinates agrees with our earlier description of
$H$ as $SU(2) \times SU(2) / U(1)$, note that the locus
determined by \hor\ is $S^3 \times S^3 = SU(2) \times SU(2)$, and the
residual $U(1)$-action \uone\ provides the correct quotient.

After removing the singularity at the origin, $X$ is topologically
$\BR_+ \times H$.  The cone metric on $X$,
\eqn\coni{ ds^2_X = dr^2 + r^2 ds^2_H \,,}
is automatically Ricci-flat due to the Einstein relation \norms.  This
metric has the same $SU(2) \times SU(2) \times U(1) = SO(4) \times
U(1)$ group of isometries as the Calabi-Yau metric discussed earlier,
and it coincides (up to normalization) with that metric.  The only 
question is how $r$ is related to $|z|$.  Since the metric is
homogeneous in $r$ of degree $2$ and homogeneous in $|z|$ of degree $4
/ 3$, we must have 
\eqn\r{r = |z|^{2/3} = \left( |z_1|^2 + |z_2|^2 + |z_3|^2 + |z_4|^2 
\right)^{1/3} \,.}  

We have discussed the geometry of $T^{1,1}$ (and the conifold) in some detail
because this example nicely illustrates the general properties of any
$H$ which preserves supersymmetry.  We will use these properties in Section
3 when we discuss supersymmetric $D3$-branes.

\subsec{The conifold gauge theory.}

Having introduced the conifold $X$, we now recall the dual superconformal
theory which arises from the infrared limit of the worldvolume theory on
coincident $D3$-branes at the conical singularity of $X$.  This theory
can be described as an infrared fixed-point of an $\CN=1$
supersymmetric, $SU(N) \times SU(N)$ gauge theory.  The gauge theory 
includes four $\CN=1$ chiral matter multiplets $A_1$, $A_2$, $B_1$, and
$B_2$.   The fields $A_1$ and $A_2$ transform in the 
$(\bk,\bkb)$ representation of the gauge group, and the fields $B_1$ and
$B_2$ transform in the $(\bkb, \bk)$ representation.
The chiral matter multiplets are coupled by a superpotential
\eqn\super{W = {\lambda \over 2} \Big( \tr \left[ A_1 B_1 A_2 B_2
\right] - \tr \left[ A_1 B_2 A_2 B_1 \right] \Big) \,,} 
where $\lambda$ is a (dimensionful) coupling constant.  

This gauge theory possesses an anomaly-free $U(1)$ $\CR$-symmetry
under which the lowest, scalar components of $A_1$, $A_2$, $B_1$, and $B_2$ 
have charge $R = 1 / 2$.  Hence the corresponding conformal dimension of
these fields at the infrared fixed-point is $\Delta = (3 / 2) \, R = 3 /
4$, and the superpotential $W$ is a marginal deformation of
the pure gauge theory at this point.  

The $\CR$-symmetry of the gauge theory corresponds to the $U(1)$
isometry arising from rotations in the $S^1$ fiber of $H$.  Note that
\uoner\ and \zs, together with the $\BZ_2$ symmetry which exchanges
the $S^2$ factors in the base of $H$, imply that the homogeneous
coordinates of $X$ also have charge $R = 1 / 2$.  Further, under a 
radial scaling on $X$, the relation \r\ implies that 
\eqn\dr{ r {\partial \over {\partial r}} z_i = {3 \over 2} z_i\,,\quad
i = 1, \ldots, 4\,,}
so the conformal dimension of the homogeneous coordinates is similarly
$\Delta = 3 / 4$.  

The $SU(2) \times SU(2)$ isometry of $H$ also appears as a
global $SU(2) \times SU(2)$ flavor symmetry in the gauge
theory.   Under this symmetry, the chiral fields (and
the corresponding homogeneous coordinates) transform as 
$({\bf 2}, {\bf 1})$ for $(A_1, A_2)$ and $({\bf 1}, {\bf
2})$ for $(B_1, B_2)$.  

However, the gauge theory in addition possesses a non-anomalous 
global $U(1)_\CB$ baryon-number symmetry which does not arise from 
an isometry of $H$.  Under this $U(1)_\CB$ symmetry, $A_1$ and $A_2$ 
have charge $+1$, and $B_1$ and $B_2$ have charge $-1$.  Not
coincidentally, these charges correspond to the charges in \uone\ of the
homogeneous coordinates under the $U(1)$ subgroup of $SU(2) \times SU(2)$
by which we quotient to obtain $T^{1,1}$.

To understand better the origin of the baryon-number symmetry, we
first note that, in analogy to the six real scalars 
$\phi_1, \ldots, \phi_6$ of the $\CN=4$ Yang-Mills theory which 
parametrize the positions of the $D3$-branes in $\BR^6$, the complex 
scalars in the chiral multiplets $A_1$, $A_2$, $B_1$, $B_2$ should parametrize
the positions of the $D3$-branes on $X$.  To show this, we observe that
the $F$-term equations which follow from \super\ imply that, modulo
descendants, 
\eqn\com{ B_1 A_i B_2 = B_2 A_i B_1\,, \quad A_1 B_j A_2 = A_2 B_j
A_1\,, \quad i,j=1,2\,.}
Consequently, if we introduce the chiral primary adjoints 
$Z_1$, $Z_2$, $Z_3$, and $Z_4$ defined by 
\eqn\zs{ Z_1 = A_1 B_1\,,\quad  Z_2 = A_2 B_2\,,\quad Z_3 = A_1
B_2\,,\quad Z_4 = A_2 B_1 \,,} 
then these operators commute modulo descendants, and their $N$ 
eigenvalues clearly satisfy the equation \con\ defining $X$.  Thus, as
we already suspect from the action of the global symmetries, the 
chiral fields $A_1$, $A_2$, $B_1$, and $B_2$ are 
matrix generalizations of the homogeneous coordinates on $X$.  The 
gauge-invariant mesonic operators made from single-traces
of products of $Z_1$, $Z_2$, $Z_3$, and $Z_4$ then describe the
locations of the $D3$-branes on $X$, and all of these operators are uncharged 
under $U(1)_\CB$.  

On the other hand, because the gauge group is $SU(N) \times SU(N)$
with chiral matter in bi-fundamental representations,
the gauge theory admits baryonic operators made from anti-symmetrizing
over the gauge indices of any $N$ bi-fundamental chiral fields.  For
instance, we have chiral primary operators 
\eqn\barya{\eqalign{ \CB^{(A)}_{i_1 \cdots i_N} &= {1 \over {N!}} \, 
\epsilon_{\alpha_1 \cdots \alpha_N} \epsilon^{\beta_1 \cdots \beta_N} 
\prod_{n=1}^N \left( A_{i_n} \right)^{\alpha_n}_{\beta_n} \,, \cr
\CB^{(B)}_{j_1 \cdots j_N} &= {1 \over {N!}} \, \epsilon^{\alpha_1
\cdots \alpha_N} \epsilon_{\beta_1 \cdots \beta_N} \prod_{n=1}^N 
\left( B_{j_n} \right)_{\alpha_n}^{\beta_n} \,, \cr}}
where the $\alpha$, $\beta$ indices are gauge indices of $SU(N) \times
SU(N)$ and the $i$, $j$ indices are flavor indices of $SU(2) \times
SU(2)$.  Note that because we anti-symmetrize over gauge-indices, 
$\CB^{(A)}_{i_1 \cdots i_N}$ and $\CB^{(B)}_{j_1 \cdots j_N}$ are 
automatically symmetric in the flavor indices.  Hence these operators 
transform under the $SU(2) \times SU(2)$ flavor symmetry as 
$({\bf N+1},{\bf 1})$ and $({\bf 1},{\bf N+1})$.  They also have
respective charges $\pm N$ under $U(1)_\CB$, justifying the
identification of this symmetry with baryon-number.  

We now introduce a schematic notation for baryonic operators, 
suppressing gauge indices and denoting 
\eqn\barysch{\CB^{(A)}_{i_1 \cdots i_N} \equiv \epsilon_1 \epsilon^2
(A_{i_1}, \cdots, A_{i_N})\,, \quad \CB^{(B)}_{j_1 \cdots j_N} \equiv
\epsilon^1 \epsilon_2 (B_{j_1},\cdots,B_{j_N}) \,.}
Here $\epsilon_1 \equiv \epsilon_{\alpha_1 \cdots \alpha_N}$ and
$\epsilon_2 \equiv \epsilon_{\beta_1 \cdots \beta_N}$ are
abbreviations for the completely anti-symmetric tensors for the 
respective $SU(N)$ factors of the gauge group.

The baryonic operators $\CB^{(A)}$ and $\CB^{(B)}$ in \barysch\ both 
have $\CR$-charge $N / 2$ and conformal dimension $\Delta = {3 \over
4} N$.  This fact makes them candidates to describe wrapped
$D3$-branes.  For if $R$ denotes the curvature scale of the 
supergravity background,
\eqn\R{ R^4 \propto g_s \alpha'^2 N\,,}
then the mass $m$ of a $D3$-brane wrapped on $H$ scales as
\eqn\m{ m \propto {{R^3} \over {g_s \alpha'^2}} \,.} 
So operators in the gauge theory which correspond to 
the \BPS\ states arising from a supersymmetric $D3$-brane wrapped on 
$H$ must be chiral primary of dimension 
\eqn\dim{ \Delta \propto m R \propto N \,.}

As first observed in \gk, the baryons $\CB^{(A)}$ and $\CB^{(B)}$ 
indeed arise from $D3$-branes wrapped on three-spheres 
in the $A$ and $B$ families described in Section 2.1.  Three-spheres
in both of these families minimize volume within their respective
homology classes, and a brane wrapped about any of these 
three-spheres is supersymmetric.  In order to determine what \BPS\
states arise from such a wrapped brane, we must quantize the bosonic
collective coordinates $(\theta_1, \phi_1)$ or 
$(\theta_2,\phi_2)$ associated to the location of the brane on the 
transverse $S^2$.  The supersymmetry generators broken by the brane
then create the additional states necessary to fill out the chiral multiplet.

The $SU(2)$ symmetry dictates that no potential is associated to the 
location of the $D3$-brane on the transverse $S^2$, but the brane does couple 
magnetically, via the Wess-Zumino term in the $DBI$ action, to the $N$ 
units of five-form flux on $H$.  (See \bhk\ for a very explicit 
demonstration of this statement.)  So the 
quantization of the collective coordinates $(\theta_1, \phi_1)$ or 
$(\theta_2,\phi_2)$ reduces to the quantization of a charged particle 
moving on the complex plane in a perpendicular
magnetic field of $N$ flux quanta.  Such a particle has $N+1$
degenerate ground-states, transforming in the $(N+1)$-dimensional
representation of $SU(2)$.  

We naturally identify these $N+1$ \BPS\ states arising from either the $A$ or
$B$ cycles as corresponding to the respective baryonic operators 
$\CB^{(A)}_{i_1 \cdots i_N}$ and $\CB^{(B)}_{j_1 \cdots j_N}$.  The
$SU(2) \times SU(2)$ flavor quantum numbers are certainly 
consistent with this identification.  Also, the baryon dimension 
$\Delta = {3 \over 4} N$ can be checked against the volume of the 
wrapped three-spheres \gk, \bhk.  

Finally, this example illustrates that the $U(1)_\CB$ baryon-number 
of the gauge theory represents a topological charge from
the perspective of the string theory, as it measures the class of the 
corresponding wrapped $D3$-brane in $\H_3$.  

\subsec{An abundance of baryons.}

Now, as observed by the authors of \bhk, we seem to have exhausted the
obvious \BPS\ states arising from supersymmetric $D3$-branes wrapped
on three-cycles in $H$.  However, we have certainly not exhausted 
all of the chiral baryonic operators in the conifold gauge theory.  

Besides the chiral fields $A_1$ and $A_2$, we have composite chiral operators
\eqn\comop{ A_{I;J} \, \equiv \, A_{i_1 \cdots i_{m+1};j_1 \cdots j_m}
\, \equiv \, A_{i_1} B_{j_1} \cdots A_{i_m} B_{j_m} A_{i_{m+1}} \,,}
which also transform as $(\bk,\bkb)$ under the gauge group and have
charge one under $U(1)_\CB$.  Here $(I,J)$ denotes the $SU(2) \times
SU(2)$ flavor representation of $A_{I;J}$.  The $F$-term equations
\com\ again imply that, modulo descendants, the operator 
$A_{i_1 \cdots i_{m+1};j_1 \cdots j_m}$ is fully symmetric on the 
flavor indices $\{ i_1,\ldots,i_{m+1} \}$ and $\{ j_1, \ldots, j_m
\}$.  Thus, for the operator in \comop, $(I,J) = ({\bf m+2},{\bf m+1})$.

Using any $N$ operators of the general form in \comop, we can form
additional chiral primary baryons $\CB^{(A)}_{R;S}$,
\eqn\newbarya{\CB^{(A)}_{R;S} = t^{I_1 \cdots I_N;J_1 \cdots J_N}_{R;S} \;
\epsilon_1 \epsilon^2 \left( A_{I_1;J_1}\,,\, \cdots\,,\, A_{I_N;J_N} \right)\,.}
Here $t^{I_1 \cdots I_N;J_1 \cdots J_N}_{R;S}$ is a Clebsch-Gordan tensor
coupling the product of flavor representations 
$(I_1, J_1) \otimes \cdots \otimes (I_N,J_N)$ to the
irreducible flavor representation $(R,S)$ of the baryon 
$\CB^{(A)}_{R;S}$.  We caution that a given flavor representation $(R,S)$
generically appears with multiplicity in this product, so the
notation $\CB^{(A)}_{R;S}$ is only schematic and usually insufficient to 
specify a given baryon.

The more general baryons $\CB^{(A)}_{R;S}$ in \newbarya\ 
transform in the same baryon-number sector as the baryon 
$\CB^{(A)}_{i_1 \cdots i_N}$.  Because baryon-number is identified as 
a topological charge measured by $\H_3$ in the string theory, the 
\BPS\ string states corresponding to the baryons $\CB^{(A)}_{R;S}$ 
must arise from supersymmetric $D3$-branes still wrapped in the class
of the $A$ cycle on $H$.  

On the other hand, the new baryonic operators have dimensions $\Delta$
strictly greater than ${3 \over 4} N$.  Since the baryons with  
$\Delta = {3 \over 4} N$ describe $D3$-branes wrapping the
minimal-volume cycles in the $A$ class, a natural interpretation of the
baryons of higher dimension is that they represent \BPS\ excitations 
of these branes.

One might hope that anyway a baryon of higher dimension could always
be factored as a product of the minimal baryon $\CB^{(A)}_{i_1 \cdots
i_N}$ and some chiral operator in the trivial baryon-number
sector.  A natural interpretation of such reducible baryons would be
as multi-particle states representing the minimally-wrapped $D3$-brane
and a topologically trivial, gravitonic excitation on $H$.

Indeed, some baryons clearly factor in this way --- but others do not. A 
baryon $\CB^{(A)}_{R;S}$ of the form \newbarya\ factors precisely when 
$R$ is the representation which is fully-symmetric on the $A$ flavor
indices.  In this case, the tensor relation 
\eqn\tensorrel{\epsilon_1 \epsilon^1 
\equiv \epsilon_{\alpha_1 \cdots \alpha_N} \,
\epsilon^{\gamma_1 \cdots \gamma_N} = \delta^{\gamma_1}_{[\alpha_1}
\cdots \delta^{\gamma_N}_{\alpha_N]}\,}
implies that $\CB^{(A)}_{R;S}$ can be factored as 
\eqn\decomp{\eqalign{ \CB^{(A)}_{R;S} \, &= \, t^{I_1 \cdots I_N;J_1 \cdots
J_N}_{R;S} \; \epsilon_1 \epsilon^1 \left( O_{K_1;J_1}\,,\,
\cdots\,,\, O_{K_N;J_N} \right) \, \cdot \,  
\epsilon_1 \epsilon^2 \left( A_{i_1}\,,\, \cdots\,,\, A_{i_n} \right) \,, \cr
&= t^{I_1 \cdots I_N;J_1 \cdots J_N}_{R;S} \; \CB^{(0)}_{K_1 \cdots
K_N;J_1 \cdots J_N} \, \cdot \, \CB^{(A)}_{i_1 \cdots i_N} \,.\cr}}

Here each $O_{K;J}$ is an adjoint chiral operator having the general form
\eqn\trivop{ O_{K;J} \, \equiv \, O_{k_1 \cdots k_m; j_1 \cdots j_m}
\, \equiv \, A_{k_1} B_{j_1} \cdots A_{k_m} B_{j_m} \,,}
again fully-symmetric on flavor indices and transforming as $({\bf
m+1},{\bf m+1})$ under the flavor symmetry.  As indicated above, 
\eqn\trivbary{\CB^{(0)}_{K_1 \cdots K_N;J_1 \cdots J_N} \equiv \epsilon_1 \epsilon^1 \left( O_{K_1;J_1}\,,\, \cdots\,,\, O_{K_N;J_N} \right) }
is a chiral primary ``baryonic'' operator in the trivial sector of 
$U(1)_\CB$ made by anti-symmetrizing the gauge indices of $N$ chiral 
operators of the form \trivop.  Also, in \decomp\ we implicitly
symmetrize over the indices in $K_1 \otimes i_1,\ldots,K_N \otimes
i_N$ to couple these representations respectively to $I_1,\ldots,I_N$.

For the case $m=0$ in \trivop, we include the constant tensor
$\delta^\gamma_\alpha$ in the general class of operators $O$.  Hence
the baryons of the form $\CB^{(0)}$ include the sub-determinants, 
\eqn\subdets{ \epsilon_{\alpha_1 \cdots \alpha_n \alpha_{n+1} \cdots
\alpha_N} \, \epsilon^{\gamma_1 \cdots \gamma_n \alpha_{n+1} \cdots 
\alpha_N} \, \prod_{k=1}^n \left( O_{I_k;J_k}
\right)^{\alpha_k}_{\gamma_k}\,, \quad 0 < n < N\,,}
as well as the trivial operator ${\bf 1}$ corresponding to $n=0$ in
\subdets.  These baryons $\CB^{(0)}$ of dimension $\Delta \sim N$,
generally identified as giant gravitons \bbns,
correspond to \BPS\ states of $D3$-branes wrapped on 
topologically trivial three-cycles in $H$, and the baryons of dimension 
$\Delta \sim 1$ correspond to multi-particle supergravity excitations 
of the metric and the four-form potential $C_4$ on $H$.

For instance, the authors of \bhk\ considered a simple class of
baryons of the form 
\eqn\simplebary{ \CB^{(A)}_{R;S} \, = \, t^{I_1 i_2 \cdots i_N}_R \; 
\epsilon_1 \epsilon^2 \left( A_{I_1;S}\,,\, A_{i_2}\,,\, \cdots\,,\, 
A_{i_N} \right) \,.}
If $R$ is the fully-symmetric representation, then $\CB^{(A)}_{R;S}$ factors as
\eqn\fac{\eqalign{ \CB^{(A)}_{R;S} \, &= \, t^{I_1 \cdots i_N}_R
\, \epsilon_1 \epsilon^1 \left( O_{K_1;S}\,,\, {\bf 1}\,,\,
\cdots\,,\, {\bf 1} \right) \, \cdot \, \epsilon_1 \epsilon^2 \left(
A_{i_1}\,,\, A_{i_2}\,,\, \cdots\,,\, A_{i_N} \right)\,, \cr
&= \,  t^{I_1 \cdots i_N}_R \; \Tr \left[ 
O_{K_1;S} \right] \, \cdot \, \CB^{(A)}_{i_1, \ldots, i_N}
\,.\cr}}
For the other representations $R$, the baryons in \simplebary\ do not
factor and represent ``new'' \BPS\ states which must be considered.
These baryons appear in a tower of evenly-spaced
$\CR$-charges ${N \over 2}, \, {N \over 2} + 1, \, {N \over 2} +
2, \ldots,$ with corresponding representations under the $SU(2)$
$B$ flavor symmetry being $S = {\bf 1}, \, {\bf 2}, \, {\bf 3},\ldots
\ldotp$  

In \bhk, the authors then perturbatively computed the spectrum of 
small fluctuations of a minimally-wrapped $D3$-brane in the $A$ class
on $H$ and found a set of \BPS\ fluctuations also having these quantum 
numbers.  They proposed that these fluctuations should correspond to 
the additional baryons in \simplebary.  But to study the general baryon 
in \newbarya, we will apply a slightly different analysis.  

\newsec{Classical supersymmetric $D3$-branes.}

The additional chiral baryons suggest additional \BPS\ $D3$-brane
states.  To account for these states, we need to find new classical,
supersymmetric $D3$-brane configurations on $H$.  So what is the
general description of a supersymmetric $D3$-brane which wraps a
three-cycle in $H$?

If we consider compactifying the type-IIB theory on a Calabi-Yau
three-fold $X$, the analogous question has a simple answer:
supersymmetric $D$-brane configurations are characterized by
holomorphy in $X$ \bbs, \ooy.  In the Freund-Rubin compactification
on $H$, might holomorphy also play a role?

Mikhailov \mik\ has answered this question affirmatively, using the
fact that compactification on $H$ is closely related to
compactification on an associated Calabi-Yau cone $X$.  As we 
shall discuss, supersymmetric $D3$-brane configurations in $H$ 
arise from holomorphic surfaces in $X$.  

First, we mention a caveat to our analysis.  In characterizing 
supersymmetric $D3$-brane configurations on $H$, for simplicity we
will ignore the degrees of freedom associated to the worldvolume
fermions and gauge field on the brane.  In particular, we assume that 
the $U(1)$ line bundle supported by the brane has a flat,
topologically trivial connection.  The latter assumption is too strong 
in general.  Whenever the $D3$-brane wraps a three-manifold in $H$ 
which has a non-trivial fundamental group $\pi_1$, we should also look 
for supersymmetric configurations with Wilson or 't Hooft loops turned on in 
the worldvolume.  Examples of these configurations occur for branes
wrapped on $S^3 / \BZ_3$ in the lens space $H = S^5 / \BZ_3$,
discussed thoroughly in \grw.

\subsec{Supersymmetric branes from holomophic surfaces.}

When $H$ preserves supersymmetry, there is always a unique Calabi-Yau
cone $X$ associated to $H$ \kehagias, \kw, \mp.  (See also \fk\ and
\bar\ for a nice mathematical discussion of these facts.)  Just as for
$T^{1,1}$, the presence of a Killing spinor on $H$ implies that the 
Einstein metric of $H$ satisfies $R_{mn} = 4 \, g_{mn}$.  Thus, if $X$ 
is the real cone over $H$, the Einstein condition implies that the
cone metric of $X$, $ds^2_X = dr^2 + r^2 \, ds^2_H$, is Ricci-flat,
and the Killing spinor on $H$ lifts to a covariantly-constant spinor on $X$.

As we know well from the \AdS/\CFT\ correspondence, the type-IIB
compactification on $M_4 \times X$, where $M_4$ denotes flat Minkowski
space, is closely related to the type-IIB compactification on $AdS_5
\times H$.  Namely, if $N$ coincident $D3$-branes fill $M_4$ and sit at
$r=0$ in $X$, then the branes warp the product metric on $M_4 \times
X$ and source $N$ units of five-form flux through $H$.  Taking the
near-horizon limit of the corresponding supergravity solution produces 
the $AdS_5 \times H$ background.

With these observations, we can easily motivate Mikhailov's
description of supersymmetric $D3$-branes in $H$.  Consider the
Euclidean theory on $\BR^4 \times X$, with $N$ $D3$-branes filling 
$\BR^4$ and one $D3$-brane wrapped on a holomorphic surface $S$ in
$X$.  We assume that $S$ intersects $H$, embedded at $r=1$ in $X$, in 
some three-manifold $\Sigma$.

This brane configuration is supersymmetric.  Just as above, we
consider the supergravity background produced by the $N$ $D3$-branes
and take the near-horizon limit.  In this limit, the radial direction of $X$
becomes a geodesic $\gamma$ in Euclidean $AdS_5$, and the probe brane
wrapped upon $S$ in $X$ becomes a probe brane wrapped on a four-manifold in
$\gamma \times H$.  The $SO(5,1)$ global symmetry of Euclidean $AdS_5$ can be
used to rotate $\gamma$ into any other geodesic in this space, and so
we can assume that $\gamma$ becomes a time-like geodesic upon
Wick-rotation back to Minkowski signature in $AdS_5$.  

This procedure produces a supersymmetric $D3$-brane wrapped on $H$,
and, running the procedure backwards, any supersymmetric $D3$-brane 
wrapped on $H$ can be lifted to a holomorphic surface $S$ in $X$.  So
we can identify the space $\CM$ which parametrizes classical, supersymmetric
$D3$-brane configurations as equivalently parametrizing holomorphic
surfaces in $X$ (which intersect the locus at $r=1$ transversely).

Because $X$ possesses many holomorphic surfaces, we have found the 
additional supersymmetric $D3$-brane configurations which we need to
get the full \BPS\ spectrum.  Yet, morally speaking, all of these
configurations are a simple generalization of the giant graviton
solution \mst\ for $H=S^5$, in that they all describe branes which carry
angular momentum around a circular orbit in $H$.

To be more explicit, we recall that the Einstein metric on $H$ always 
possesses a $U(1)$ isometry, generated by a Killing vector 
${\partial \over {\partial \phi}}$, which
corresponds to the $\CR$-symmetry of the $\CN=1$ superconformal
algebra.  Provided that this $U(1)$-action is regular, meaning that 
its orbits are compact and of uniform length, as we now 
assume, then $H$ is a principal $U(1)$-bundle over a
positively-curved K\"ahler-Einstein manifold\foot{If these orbits are 
only compact, then $V$ is an orbifold \bg.} $V$.

Let $\tau$ be a unit-speed parameter along the geodesic $\gamma$ in
$AdS_5$, and let $\phi$ be a unit-speed parameter along the $U(1)$
fiber over $V$.  At the instant $\tau = 0$, corresponding to $r=1$ in
$X$, the $D3$-brane configuration on $H$ is described by the
three-manifold $\Sigma$.  Holomorphy of $S$ in $X$ then implies that the
$D3$-brane configuration in $\gamma \times H$ depends only on $\tau$
and $\phi$ through the combination $\tau + \phi$.  As a result, the brane is
described just by translating $\Sigma$ at unit-speed around the $U(1)$
fiber over $V$.

One great advantage of our characterization of supersymmetric
$D3$-brane configurations via holomorpy is that it does
not require explicit knowledge of the Einstein metric on $H$.  We must
only know the appropriate complex structure on $X$, which is usually
obvious.  Also, at least in the examples $H = T^{1,1}$ and $H = S^5$,
we can very simply parametrize the relevant holomorphic surfaces
in $X$, and thus describe $\CM$ directly.

\subsec{Holomorphic surfaces in the conifold.}

Let $H = T^{1,1}$ and $X$ be the conifold.  Holomorphic surfaces 
$S$ in $X$ are easy to describe.  Because we defined $X$ through 
an embedding in $\BC^4$, a very obvious set of holomorphic surfaces 
arise from the intersection of hypersurfaces in $\BC^4$ with $X$.  The 
general hypersurface is simply determined as the vanishing 
locus of a polynomial $P$ in the affine coordinates
$(z_1,z_2,z_3,z_4)$ of $\BC^4$.

However, the naive approach of using hypersurfaces in $\BC^4$ to
describe holomorphic surfaces in $X$ will not quite work.  Some 
polynomials, for instance $P = z_1 z_2 - z_3 z_4$, restrict to zero on $X$ and
therefore do not actually determine a holomorphic surface in $X$.
Furthermore, if we were to attempt to parametrize surfaces in $X$ by
parametrizing polynomials in the affine coordinates of $\BC^4$, we
would have to face the issue that, on $X$, these polynomials are only 
well-defined up to the addition of the polynomials which vanish 
identically on $X$.  Finally, any holomorphic surface in $X$ which
does arise from a hypersurface in $\BC^4$ describes a $D3$-brane which
is wrapped in the trivial homology class of $H$, and we are most
interested in branes wrapped in the (non-trivial) $A$ class of $H$.

These difficulties can all be overcome if we use the homogeneous
coordinates $A_1$, $A_2$, $B_1$, and $B_2$ on $X$.  If $P$ is a
polynomial in the these coordinates which transforms homogeneously
under the $U(1)$-action in \uone, then the vanishing locus of $P$ is a 
well-defined holomorphic surface in $X$.  Unlike the affine
coordinates of $\BC^4$, though, there are no algebraic relations among 
the homogeneous coordinates on $X$.  

In addition, polynomials $P$ which are charged under the $U(1)$-action
in \uone\ can be used to describe topologically non-trivial wrappings
of branes on $H$.  In fact, the charge of $P$ under this $U(1)$ 
corresponds to the class of the wrapped three-cycle in 
$\H_3(H) = \BZ$.  For instance, if $P$ has charge zero, then the 
relations \zs\ imply that $P$ arises from the restriction of a
hypersurface in $\BC^4$.  As we have mentioned, these hypersurfaces
all describe trivially wrapped $D3$-branes on $H$.  

To describe $D3$-branes on $H$ which wrap in the $A$ class of $H$, we
consider polynomials $P(A_1,A_2,B_1,B_2)$ which have charge one.  Such
a polynomial has the form
\eqn\gen{\eqalign{ P(A_1,A_2,B_1,B_2) &= c_1 \, A_1 + \; c_2 \, A_2
\cr &+ \; c_{11;1} \, A_1^2 B_1 \quad + \; c_{12;1} \, 
A_1 A_2 B_1 \quad + \; c_{22;1} \, A_2^2 B_1 \cr &+ \;
c_{11;2} \, A_1^2 B_2 \quad + \; c_{12;2} \, A_1 A_2 B_2
\quad + \; c_{22;2} \, A_2^2 B_2 \cr
&+ \; \left( \hbox{terms of higher degree in } A_1, A_2 \right) \,.}}
Here the $c$'s are just arbitrary complex coefficients for each
monomial in $P$.  To write this expression more concisely, we
introduce multi-indices $I$ and $J$ so that
\eqn\geng{ P(A_1,A_2,B_1,B_2) \, = \,
\sum_{I,J} \; c_{I;J} \, A^I B^J \,.}

We can consider the coefficients $c_{I;J}$ as complex
coordinates $(c_1, \, c_2, \, c_{11;1}, \, \cdots)$ on the 
infinite-dimensional, complex vector space of polynomials $P$ which
have charge one.  Each point in this vector space, other than the
origin, determines a holomorphic surface in $X$.  And, modulo the
trivial fact that $P$ and $\lambda P$, for any $\lambda \in
\BC^\times$, vanish on the same locus in $X$, the points in this
vector space describe distinct surfaces.  Thus, we identify the
classical configuration space $\CM$ of supersymmetric branes wrapped
in the $A$ class on $H$ as $\BC \BP^\infty$, with homogeneous coordinates 
$[c_1: \, c_2: \, c_{11;1}: \, \cdots]\,.$

We can fairly easily visualize what branes some of these polynomials
describe.  As an example, let us consider a linear polynomial 
\eqn\linpoly{ P = c_1 \, A_1 + \; c_2 \, A_2\,.}
Both terms in $P$ have $\CR$-charge $1/2$, so the zero locus of $P$ is
fixed under the $U(1)$ isometry of $H$, meaning that the brane described by
$P$ wraps a fixed three-manifold $\Sigma$ independent of time.  Since 
$A_1$ and $A_2$ descend to homogeneous coordinates on one
$\BC \BP^1$ factor of the $\BC \BP^1 \times \BC \BP^1$ base of $H$, 
$P$ vanishes over a point, parametrized by $c_1$ and $c_2$, on this 
$\BC \BP^1$.  So this $D3$-brane configuration corresponds to one
of the well-known supersymmetric $D3$-branes which wrap a minimal-volume $S^3$
in the $A$ class.

One the other hand, when $P$ is of the general form in \gen, the
various terms have different $\CR$-charges and $P$ transforms
inhomogeneously under the $\CR$-symmetry.  Hence the general 
$D3$-brane configuration corresponds to some $\Sigma$ orbiting the 
$U(1)$ fiber over $\BC \BP^1 \times \BC \BP^1$.  To obtain a more
explicit description of $\Sigma$, we would actually have to solve the 
equation $P = 0$ on $H$ --- but of course the important
issue is not so much the particular description of a given
brane configuration, but the global description of the entire 
family $\CM$ of configurations.  

We can equally well consider $D3$-branes wrapped in other 
homology classes, corresponding to polynomials of other charges, and
the above comments remain unaltered.  $\CM$ is always an
infinite-dimensional projective-space whose homogeneous coordinates
correspond to the coefficients in the homogeneous polynomials of fixed
charge --- except for a subtlety relating to the trivially-wrapped 
$D3$-branes.  We will explain this point as we consider the 
example $H = S^5$, corresponding to $X = \BC^3$.

\subsec{Holomorphic surfaces in $\BC^3$.}

Let $(z_1,z_2,z_3)$ be affine coordinates on $X = \BC^3$.  The relevant $U(1)$
isometry of $H = S^5$ corresponds to
\eqn\uoners{ (z_1,z_2,z_3) \rightarrow (e^{i \alpha} \, z_1, e^{i
\alpha} \, z_2, e^{i \alpha} \, z_3)\,,\quad \alpha \in [0,2 \pi)\,.}
The space of orbits for this $U(1)$-action on $H$ is just $\BC \BP^2$.

Any holomorphic surface in $X$ can be presented as a hypersurface
determined by the zero locus of a polynomial $P(z_1,z_2,z_3)$.  For
concreteness, we parametrize the polynomial $P$ as 
\eqn\poly{\eqalign{ P &= c + c_i \, z_i + c_{ij}\, z_i z_j +
c_{ijk} \, z_i z_j z_k + \cdots\,, \cr
&= \sum_I \, c_I \, z^I \,,\cr}}
where $c,\, c_i,\, c_{ij},\, c_{ijk},\, \ldots$, are complex 
coefficients and in the second line of \poly\ we introduce a
multi-index $I$ for brevity.  As for the conifold, we regard these
coefficients as homogeneous coordinates
$[c:c_i:c_{ij}:c_{ijk}:\cdots]$ on a $\BC \BP^\infty$ parameter 
space.  

However, the new ingredient in this example is the presence of a
constant term $c$ in $P$.  The associated point $[1:0:0:0:\ldots]$ in
$\BC \BP^\infty$ corresponds to the nowhere-vanishing, constant
polynomial $P = 1$.  The constant polynomial of course does not
correspond to any hypersurface in $X$ nor a $D3$-brane configuration
on $H$.

Further, the general construction of a supersymmetric $D3$-brane 
configuration on $H$ requires that the holomorphic surface $S$
determined by $P$ intersects $H$ transversely at $r=1$.  So points 
near $[1:0:0:0:\ldots]$, for which $c \gg c_i, c_{ij}, c_{ijk},
\ldots$, do not determine surfaces which intersect $H$ and thus do not 
determine any $D3$-brane configurations either.   

In this case, the classical configuration space $\CM$ does not consist 
of the full $\BC \BP^\infty$ parameter space, but only an open subset 
of this space away from the bad point at $[1:0:0:0:\ldots]$.
Nevertheless, we will see that quantum mechanically this example is
much like the simpler case for which $\CM = \BC \BP^\infty$.

Yet the fact that we include the constant term in the general
polynomial is important.  A simple example to consider is an affine
polynomial of the form $P = z_1 - c$.  For $|c|<1$, this polynomial 
defines a hyperplane in $X$ which intersects $H$ in an $S^3$.  As long
as $c \neq 0$, the $U(1)$-action in \uoners\ translates this $S^3$ around
a circle in the $H$.  So the affine polynomials describe the giant
gravitons, and $c$ controls the size of the wrapped $S^3$.

Another simple case to consider is a polynomial of arbitrary degree $n$
but which depends on only a single coordinate such as $z_1$.  In this 
case, we can factor the polynomial as 
\eqn\factp{ P = \prod_{i=1}^n (z_1 - e_i)\,.}
Provided that all of the roots $e_i$ satisfy $|e_i|<1$, this 
polynomial represents a configuration of $n$ giant gravitons of
various sizes on $H$.  

This example is important, as it illustrates the general principle
that reducible surfaces, corresponding to factorizable polynomials,
represent configurations of multiple branes.  So when we quantize the
space $\CM$, which includes both irreducible and reducible surfaces,
we automatically ``second quantize'' the $D3$-brane.  We shall discuss some
additional aspects of the relation of reducible surfaces to
multi-brane states at the end of Section 4.

\subsec{A technical aside.}

The following is a somewhat technical aside, not necessary for reading 
the rest of the paper but possibly useful if one wishes to generalize
the analysis which we performed for non-trivial wrappings in $H =
T^{1,1}$ to other examples.  In writing global expressions for the 
holomorphic surfaces in the conifold $X$, we naturally used the 
global homogeneous coordinates $(A_1,A_2,B_1,B_2)$.  The
existence of such homogeneous coordinates is related to the fact that 
$X$ is a toric variety.  But for the general $X$, global homogeneous 
coordinates need not exist.  Nevertheless, we can still very easily 
describe the classical configuration space $\CM$.

To start, we will present a slightly different characterization of
supersymmetric $D3$-branes which wrap three-cycles in $H$.  $H$ can be
described as a principal $U(1)$-bundle over a K\"ahler-Einstein
manifold $V$.  This bundle is, in a suitable sense, holomorphic.  If 
$A$ is the unitary connection on this bundle arising from the metric
of $H$, so that
\eqn\hmet{ ds^2_H = (d \phi + i A)^2 + ds^2_V \,,}
then the curvature $F = d A$ is proportional to the K\"ahler form
$\omega$ of $V$.  In particular, $F$ is of type $(1,1)$ with respect
to the complex structure of $V$.  Hence, this $U(1)$ bundle over $V$
can always be considered to arise as the unit-circle bundle of some
holomorphic line bundle over $V$.

A $D3$-brane configuration on $H$ is described by a four-manifold in
$\gamma \times H$.  By the above remarks, we can regard $\gamma \times
H$ (in Euclidean signature) as the total space $E$ of a 
holomorphic $\BC^\times$-bundle over $V$.  The holomorphic
fiber coordinate is just $\tau_E + i \phi$, where $\tau_E = i \, \tau$.

Supersymmetric $D3$-brane configurations then correspond to 
holomorphic surfaces in $E$ with one holomorphic tangent vector along 
the $\BC^\times$ fiber, i.e. any holomorphic surface in $E$ other than 
$V$ itself.  One can verify this statement directly using the 
kappa-symmetric $D3$-brane worldvolume action \cgnw, \aps, \cgnsw,
\bt, \apsa\ as generally illustrated in \bkop.  In this analysis, the 
condition for supersymmetry is expressed geometrically as a
generalized calibration condition \harvey, \gut\ which is satisfied by these 
holomorphic surfaces in $E$.

Now let $D$ be a divisor on $V$.  We will consider the supersymmetric 
$D3$-branes on $H$ such that the projection of the 
$D3$-brane worldvolume to $V$ is a divisor in the linear equivalence class of
$D$.  Let $\CO(D)$ be the line-bundle on $V$ associated to $D$, and
let $\pi:E \rightarrow V$ denote the projection, so that $\pi^* \CO(D)$
is a line-bundle on $E$.  Then the vanishing loci of the 
global holomorphic sections of $\pi^* \CO(D)$ naturally determine holomorphic
surfaces in $E$ representing this class of supersymmetric $D3$-brane 
configurations.  The classical configuration space $\CM$ is just the
projective space associated to the vector space of global holomorphic sections 
$\H^0 \! \left( E,\, \pi^* \CO(D) \right)$.  

\newsec{Quantum supersymmetric $D3$-branes.}

We finally turn to quantizing the classical configuration space
$\CM$.  Because we are only considering the Hilbert space of \BPS\
states in the brane spectrum, we do not really need any more detailed
information about $\CM$ other than its complex structure, which we have
already described.  In particular, although we assume that $\CM$
possesses some K\"ahler metric, the particular form of the metric is 
mostly irrelevant in the following analysis --- a very fortunate thing, 
since we do not know anything about it.  The philosophy we apply falls 
under the rubric of geometric quantization, for which a general
reference is \woodhouse.

For concreteness, we now discuss the examples $H = T^{1,1}$ and $H
= S^5$, for which we can compare our results to the dual baryon spectrum.

\subsec{Quantum $D3$-branes on $H = T^{1,1}$.}

In this case, the classical configuration space $\CM$ which describes 
the supersymmetric $D3$-branes wrapped in the $A$ class of $H$ is  
$\BC \BP^\infty$, with homogeneous coordinates 
$[c_1: \, c_2: \, c_{11;1}: \, \cdots: c_{I;J}: \, \cdots]$ corresponding
to the coefficients of the polynomial in \gen.  We can think of 
the $D3$-brane as a particle moving on $\CM$, and so
the $D3$-brane phase space is the cotangent bundle $T^* \CM$.   The 
$D3$-brane wavefunction $\Psi$ takes values in some holomorphic 
line-bundle $\CL$ over $\CM$, and the Hamiltonian which acts on $\Psi$ is
the Laplacian acting on sections of $\CL$.  The \BPS\ wavefunctions
are the global holomorphic sections of $\CL$, and the
vector space of these sections is the \BPS\ Hilbert space.

To proceed, we only have to describe the line-bundle $\CL$, which is
easy since $\CL$ is determined by its degree.  To compute the degree
of $\CL$, we consider any $\BC \BP^1$ subspace of $\CM$ and a
one-parameter family of classical $D3$-brane configurations
corresponding to a curve $\CC$ in this subspace.  If the $D3$-brane
configuration changes with time, so that the $D3$-brane traverses the
curve $\CC$ in $\CM$, then $\Psi$ picks up a phase because the
$D3$-brane couples to the background $RR$ four-form potential $C_4$.
This phase determines the degree.

Let $s$ be a parameter along $\CC$, and $A(s)$ the corresponding
family of three-manifolds in the class of $A$.  Finally let $D$ be a
disc in $\BC \BP^1$ bounding $\CC$.  If we take the curve $\CC$ in
$\BC \BP^1$ to be large, $D$ can be taken to cover the full 
$\BC \BP^1$ subspace.  The corresponding two-parameter family of
$D3$-branes sweeps out all of $H$, and the phase which $\Psi$
acquires when the $D3$-brane traverses the curve $\CC$ is 
\eqn\phase{ 2 \pi i \, \oint_\CC d s \, \int_{A(s)} \! C_4 \, = \, 2
\pi i \, \int_{H} F_5 \, = \, 2 \pi i \, N \,.}

So $\CL$ has degree $N$, i.e. $\CL = \CO(N)$.  In terms of the
homogeneous coordinates $[c_1: \, c_2: \, c_{11;1}: \, \cdots:
c_{I;J}: \, \cdots ]$ of $\CM$, the space of global holomorphic 
sections of $\CL$ consists of the degree $N$ polynomials in these 
coordinates.  Thus, the Hilbert space of \BPS\ states is spanned by 
states of the form
\eqn\monos{ | c_{I_1;J_1} \cdots c_{I_N;J_N} \rangle \,.}

At this point, we see that we really need not worry about any
subtleties in dealing with holomorphic bundles on infinite-dimensional
complex manifolds.  Since $\CL$ has only finite degree, any given 
wavefunction $\Psi$ in the \BPS\ sector varies non-trivially only over 
some finite-dimensional subspace of $\CM$.  If we were to filter the 
polynomials in \gen\ by their degree in $A_1$ and $A_2$, we would be 
perfectly justified to consider inductively the corresponding
filtration of $\CM$ by finite-dimensional subspaces $\BC \BP^1 
\subset \BC \BP^7 \subset \cdots \subset \CM $ of successively
increasing dimension.

We now describe the correspondence between the \BPS\ states in the
$D3$-brane Hilbert space and the baryonic operators in the conifold gauge
theory.  Note that we have an obvious association between the
homogeneous coordinates of $\CM$ and the chiral operators of 
$U(1)_\CB$ charge one in the gauge theory,
\eqn\assc{ c_{i_1 \ldots i_{m+1};j_1 \ldots j_m} \; \longleftrightarrow
\; A_{i_1} B_{j_1} A_{i_2} \cdots B_{j_m} A_{i_{m+1}} \,.}
We may of course permute the elements of $\{i_1 \ldots i_{m+1}\}$, 
$\{j_1 \ldots j_m\}$ as we like in the case of the $c$'s,
and the $F$-term equations ensure that the same is true for the
chiral operators.  So we propose that the correspondence
between suitably normalized \BPS\ states and baryons is that
\eqn\proposal{ | c_{I_1;J_1} \cdots c_{I_N;J_N} \rangle \;
\longleftrightarrow \; \epsilon_1 \epsilon^2 \left( A_{I_1;J_1}\,,\,
\cdots\,,\, A_{I_N;J_N} \right) \,.}

Note that both the \BPS\ state and the baryon in \proposal\ are fully 
symmetric in the indices $\{1,2,\ldots,N\}$.  This property is a
trivial property of degree $N$ monomials and arises due to the
anti-symmetrization of gauge indices in the baryon.  Also, the global
homogeneous coordinates $c_{I;J}$ transform under the induced action 
of the $SU(2) \times SU(2) \times U(1)$ isometry of $X$ and the 
radial scaling of $X$ exactly as the associated chiral matter fields 
$A_{I;J}$ transform under the corresponding symmetries of the gauge 
theory.  So the \BPS\ states and baryons in \proposal\ have manifestly 
the same global quantum numbers.  Finally, this proposal naturally 
generalizes the well-known correspondence of the baryonic operators 
$\CB^{(A)}_{i_1 \cdots i_N}$ of least conformal dimension to the \BPS\ 
states $|c_{i_1} \cdots c_{i_N} \rangle$ which arise from sections 
of $\CO(N)$ over the $\BC \BP^1$ subspace of $\CM$ corresponding 
to the linear polynomials in \linpoly.

This quantization procedure directly extends to $D3$-branes
wrapped in the classes, such as $2 A, 3 A, \ldots$, of higher degree in $\H_3$.
However, as observed in \bhk, the gauge theory does not seem to
possess any new baryons in these sectors beyond those arising from
products of baryons in the basic $A$ and $B$ sectors.  At this point,
we can only venture to speculate that the \BPS\ states which would
naively arise from branes wrapped in classes of higher degree on $H$
are susceptible to decay into multi-particle \BPS\ states arising from 
the branes wrapped in the generating $A$ and $B$ classes.

\subsec{Quantum $D3$-branes on $H = S^5$.}

In this example, the classical configuration space $\CM$ consists only
of an open subset away from the point $[1:0:0:0:\ldots]$ of the 
$\BC \BP^\infty$ parameter space with homogeneous coordinates 
$[c:c_i:c_{ij}:c_{ijk}:\cdots:c_I:\cdots]$.  The same argument as for
$H = T^{1,1}$ implies that the line-bundle $\CL$ over $\CM$ has
degree $N$ and, for wavefunctions which vary non-trivially only over
the projective subspace $\CM_0 \subset \CM$ at $c = 0$, holomorphic 
sections of $\CL$ correspond to degree $N$ polynomials in 
$c_i, c_{ij}, c_{ijk}, \ldots \ldotp$

One might worry that, in this case, the non-compactness associated
to the coordinate $c$ would lead to many additional wavefunctions, of
arbitrary degree in $c$.  However, the essential point is that $\CM$
does contain the projective subspace $\CM_0$ and can be considered
here as the total space of the line-bundle $\CO(1)$ over $\CM_0$.  The
twisting in the fiber of this bundle ensures that a section $\Psi$ of
$\CL$, having at worst a pole of degree $N$ on $\CM$, is still a
polynomial of at most degree $N$ in $c$, just as for the homogeneous 
coordinates on the base $\CM_0$.

Thus, despite the fact that the nature of $\CM$ is slightly different
depending upon whether the $D3$-brane wraps a topologically
non-trivial or a trivial three-cycle, in both cases the \BPS\
wavefunctions are degree $N$ polynomials in the global homogeneous
coordinates.  In the case $H = S^5$, the \BPS\ states are now linear 
combinations of the states 
\eqn\monoggs{ | c_{I_1} \cdots c_{I_N} \rangle \,.}

To describe the corresponding operators in the $\CN=4$ Yang-Mills
theory, we decompose the $\CN=4$ Yang-Mills multiplet into an 
$\CN=1$ vector multiplet and three $\CN=1$ chiral multiplets 
$Z_1$, $Z_2$, $Z_3$, whose scalar components consist of complex
combinations of the six Yang-Mills scalars $\phi_1,\ldots,\phi_6$,
\eqn\cmplx{ z_1 = {1 \over {\sqrt{2}}} (\phi_1 + i \phi_2)\,,\quad z_2
= {1 \over {\sqrt{2}}} (\phi_3 + i \phi_4)\,, \quad z_3 = 
{1 \over {\sqrt{2}}} (\phi_5 + i \phi_6) \,.} 

We then associate the coordinates $c_I$ with chiral operators as
\eqn\asscgg{ c_I \, \longleftrightarrow \, Z_I \,.}
The cubic superpotential that appears when the $\CN=4$ theory is
decomposed into $\CN=1$ representations again ensures that both the
left-hand and the right-hand sides of \asscgg\ are symmetric in the
multi-indices $I$.  The novel feature here is that the constant term
$c$ in \poly\ is associated to the 
trivial chiral operator ${\bf 1}$ in the Yang-Mills theory.

As for $H = T^{1,1}$, we propose the natural correspondence between
\BPS\ states and baryonic operators,
\eqn\proposalgg{ | c_{I_1} \cdots c_{I_N} \rangle \,
\longleftrightarrow \, \epsilon_1 \epsilon^1 \left(
Z_{I_1}\,,\, \cdots\,,\, Z_{I_N} \right) \,.}
For instance, we see that the sub-determinant operators  
\eqn\gg{ \epsilon_1 \epsilon^1 
\left( \overbrace{1,\ldots,1}^{k},\overbrace{Z_1,\ldots,Z_1}^ {N-k}
\right)}
correspond to \BPS\ states which arise from the affine polynomials $P =
c + c_1 \, z_1$ classically describing giant gravitons.  This
observation seems to provide a nice conceptual basis for the proposal
of \bbns.

\subsec{Multi-particle states and reducible surfaces.}

From our description of $\CM$, we see that the generic point of $\CM$
describes an irreducible surface in $X$, but there is a closed subset
$\CM_R$ which consists of reducible surfaces.  Only
the irreducible surfaces classically describe single branes, and the
reducible surfaces describe configurations of multiple branes.  So
when we quantize $\CM$, some of the wavefunctions we obtain describe 
single-particle states, and others must describe multi-particle states.

But which of our \BPS\ states are the single-particle states, and
which are the multi-particle states?  The answer to this question must
involve the behavior of the K\"ahler metric on $\CM$, since only the
single-particle wavefunctions are normalizable in $L^2$ on $\CM$.  A
reasonable hypothesis, based on the classical observations above, is
that this metric is such that the reducible locus $\CM_R$ lies at
infinite distance from all other points in $\CM - \CM_R$.  

If this guess is correct, then any single-particle wavefunction on
$\CM$, being normalizable, would necessarily vanish over the locus
$\CM_R$, as the classical reasoning suggests.  Conversely,
wavefunctions on $\CM$ not vanishing over $\CM_R$, i.e. 
extending to ``infinity'', would naturally be interpretable as 
non-normalizable, multi-particle states.

To describe how this idea can be interpreted in a simple case, for
which $X$ is the conifold, we consider the locus in $\CM_R$
corresponding to factorizable polynomials of the form 
\eqn\redpoly{P(A_1,A_2,B_1,B_2) = \left( a_1 \, A_1 + a_2 \, A_2
\right) \cdot \left( \sum_{I;J} \; b_{I;J} \, A^I B^J \right) \,,}
where the second factor is a general polynomial of charge zero.  The 
coefficients $[a_1:a_2]$ and $[b:b_{1;1}:b_{2;1}:\cdots:b_{I;J}:\cdots]$ 
determine homogeneous coordinates on a $\BC \BP^1 \times \BC
\BP^\infty$ family\foot{More precisely, we should remove the point of
the $\BC \BP^\infty$ factor corresponding to the constant polynomial,
so that we only consider non-trivial factorizations of $P$.} of such
reducible surfaces in $X$.  There is a natural map 
$f:\BC \BP^1 \times \BC \BP^\infty \rightarrow \CM$
from this family of reducible surfaces into the general space of
holomorphic surfaces $\CM$.  This map just corresponds to multiplying 
out the coefficients in \redpoly, so in terms of the homogeneous 
coordinates $c_{I;J}$ of $\CM$, $f$ is expressed by 
\eqn\map{ c_{\, i_1 i_2 \cdots i_{m+1} ; j_1 \cdots j_m} = a_{(i_1}
\cdot b_{i_2 \cdots i_{m+1});j_1 \cdots j_m}\,,}
where we symmetrize over the indices $\{i_1,\ldots,i_{m+1}\}$
appearing on the right-hand side of \map.  Under $f$, sections of $\CO(N)$ on 
$\CM$ pull-back to sections of $\CO(N) \otimes \CO(N)$ on 
$\BC \BP^1 \times \BC \BP^\infty$.
 
We now consider a general element of the Hilbert space of the form 
\eqn\genstate{ t^{I_1 \cdots I_N; J_1 \cdots J_N}_{R;S}
| c_{I_1;J_1} \cdots c_{I_N;J_N} \rangle \,,}
where $t^{I_1 \cdots I_N; J_1 \cdots J_N}_{R;S}$ is the same tensor used to
make the general baryon $\CB^{(A)}_{R;S}$ in \newbarya.  Upon pull-back,
\eqn\plback{ f^* \left(t^{I_1 \cdots I_N; J_1 \cdots J_N}_{R;S} |
c_{I_1;J_1} \cdots c_{I_N;J_N} \rangle \right) \, = \, 
t^{I_1 \cdots I_N; J_1 \cdots J_N}_{R;S}
| a_{i_1} \, \cdots \, a_{i_N} \rangle \otimes | b_{K_1;J_1} \, \cdots \,
b_{K_N;J_N} \rangle \,,}
where our notation is as for the factorizable baryon in \decomp.

We now make essentially the same observation which we made for the baryons in
Section 2.3.  The tensor product of states on the right-hand side of
\plback\ necessarily transforms in the representation which is
fully-symmetric on $SU(2)$ $A$ flavor indices --- so the pull-back of 
the general state in \genstate\ to this reducible locus vanishes unless $R$ is
the fully-symmetric representation.  

Thus our guess about the metric on $\CM$ implies that a state 
of the form \genstate\ for which $R$ is the fully-symmetric
representation must be a multi-particle state.  As we have seen in 
Section 2.3, the corresponding baryon always factors into a product of
the minimal baryon $\CB^{(A)}_{i_1 \cdots i_N}$ and some other baryon 
$\CB^{(0)}_{K_1 \cdots K_N;J_1 \cdots J_N}$.  Under our general
proposal relating \BPS\ states and baryons, we naturally identify this
product of baryons as corresponding to the tensor product of states in
\plback. 

\bigbreak\bigskip\bigskip\centerline{{\bf Acknowledgements}}\nobreak
It is a pleasure for me to thank Aaron Bergman, Chris Herzog, Peter Ouyang, 
and Natalia Saulina for stimulating conversations on these
matters.  I would like especially to thank my advisor, Edward Witten,
for initially suggesting this project to me and then for providing a
tremendous amount of help and essential advice.
 
The author is supported by an NSF Graduate Fellowship and 
under NSF Grant PHY-9802484.

\listrefs

\end